\documentclass[12pt]{article}
\usepackage{graphicx}

\newcommand\pubnumber{DPF2013-114}
\newcommand\pubdate{\today}

\def\napoli{University of Virginia, Charlottesville, Virginia
22904, USA\\
Fermi National Accelerator Laboratory, Batavia, Illinois 60510, USA}

\def\support{\footnote{Work supported by the DOE Office of Science.}}

\def\Title#1{\begin{center} {\Large #1 } \end{center}}
\def\Author#1{\begin{center}{ \sc #1} \end{center}}
\def\Address#1{\begin{center}{ \it #1} \end{center}}

\newcommand\pubblock{\rightline{\begin{tabular}{l} \pubnumber\\
         \pubdate  \end{tabular}}}
\newenvironment{Abstract}{\begin{quotation}  }{\end{quotation}}
\newenvironment{Presented}{\begin{quotation} \begin{center} 
             PRESENTED AT\end{center}\bigskip 
      \begin{center}\begin{large}}{\end{large}\end{center} \end{quotation}}

\def\beq{\begin{equation}}
\def\eeq#1{\label{#1}\end{equation}}
\def\eeqn{\end{equation}}

\def\beqa{\begin{eqnarray}}
\def\eeqa#1{\label{#1}\end{eqnarray}}
\def\eeqan{\end{eqnarray}}

\let\bar=\overbar

\def\Dslash{\not{\hbox{\kern-4pt $D$}}}
\def\dslash{\not{\hbox{\kern-2pt $\del$}}}

\def\msb{{\bar{\ssstyle M \kern -1pt S}}}

\begin{document}
\begin{titlepage}
\pubblock

\vfill
\Title{Design considerations for the cosmic-ray-veto system of the Mu2e experiment at Fermilab}
\vfill
\Author{ Craig Group and Yuri Oksuzian\support \\
(on behalf of the Mu2e Collaboration)}
\Address{\napoli}
\vfill
\begin{Abstract}
Since the discovery of the muon, particle physicists have carried out a series of experiments aimed at measuring flavor violation in charged-lepton interactions. To date, no such violation has been observed. The Mu2e experiment at Fermilab will search for the charged lepton flavor violating process of coherent muon-to-electron conversion in the presence of a nucleus with a sensitivity four orders of magnitude beyond current limits. The experiment will have a single event sensitivity of about $2.5\times10^{-17}$ while limiting the total background to less than 0.5 events. One potential background is due to cosmic-ray muons producing an electron that is indistinguishable from signal within the Mu2e apparatus. The cosmic-ray-veto system of the Mu2e experiment is tasked with vetoing cosmic-ray-induced backgrounds with high efficiency without inducing significant dead time and while operating in a high-intensity environment. In this note the challenges related to the high-radiation environment that influence the design of the cosmic-ray-veto system will be discussed.
\end{Abstract}

\vfill
\begin{Presented}
DPF 2013\\
The Meeting of the American Physical Society\\
Division of Particles and Fields\\
Santa Cruz, California, August 13--17, 2013\\
\end{Presented}
\vfill
\end{titlepage}
\def\thefootnote{\fnsymbol{footnote}}
\setcounter{footnote}{0}

\section{Introduction}

     The Mu2e experiment will search for the conversion of a muon into an
  electron in the presence of a nucleus~\cite{Mu2eCDR} at a
  sensitivity of about four orders of magnitude beyond the best
  current limits.  The observation of this process would be a major
  discovery, signaling the existence of charged-lepton-flavor
  violation far beyond what is expected from current standard theory.
  A non-observation would be equally interesting as it would place
  strong limits on theory and would exclude large regions of parameter space
  for leading theories of beyond standard model physics.

\begin{figure}[htb]
\centering
\includegraphics[width=6in]{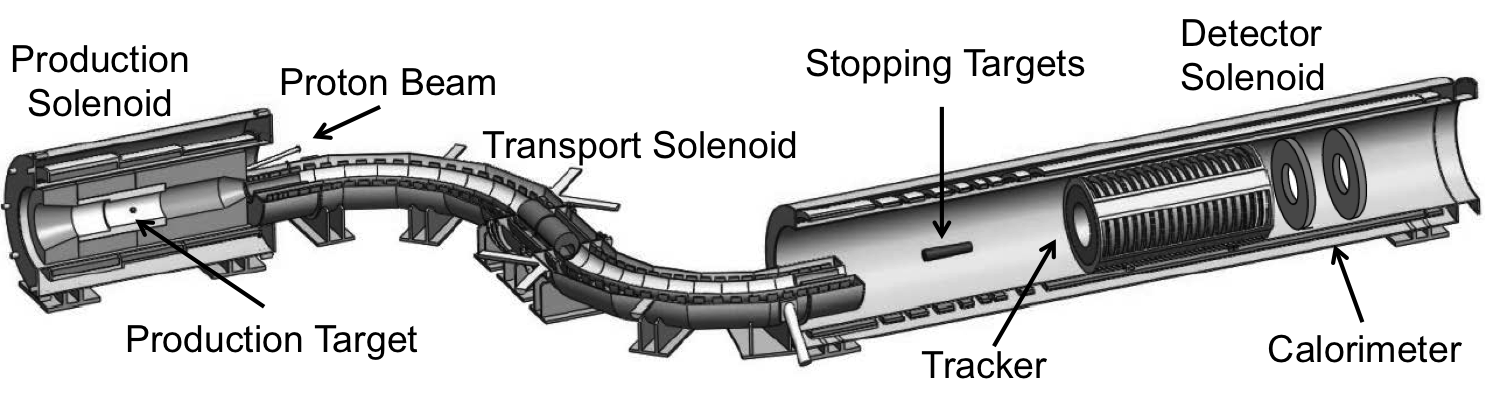}
\caption{  The Mu2e experimental setup. The pulsed proton beam enters the production solenoid as shown.  The muons produced are captured by the Production Solenoid and transported through the S-shaped bend of the Transport Solenoid to the stopping target.  Electrons produced in the stopping target are captured by the magnetic field in the Detector Solenoid and transported through the Tracker where the momentum is measured.  The electrons then strike the Electromagnetic Calorimeter, which provides an independent measurement.  A cosmic ray veto system surrounds the Detector Solenoid region and part of the Transport Solenoid, but is not shown in the figure.}
\label{fig:Mu2e}
\end{figure}

\begin{figure}[hbt]
  \centering
  \includegraphics[width=2.8in]{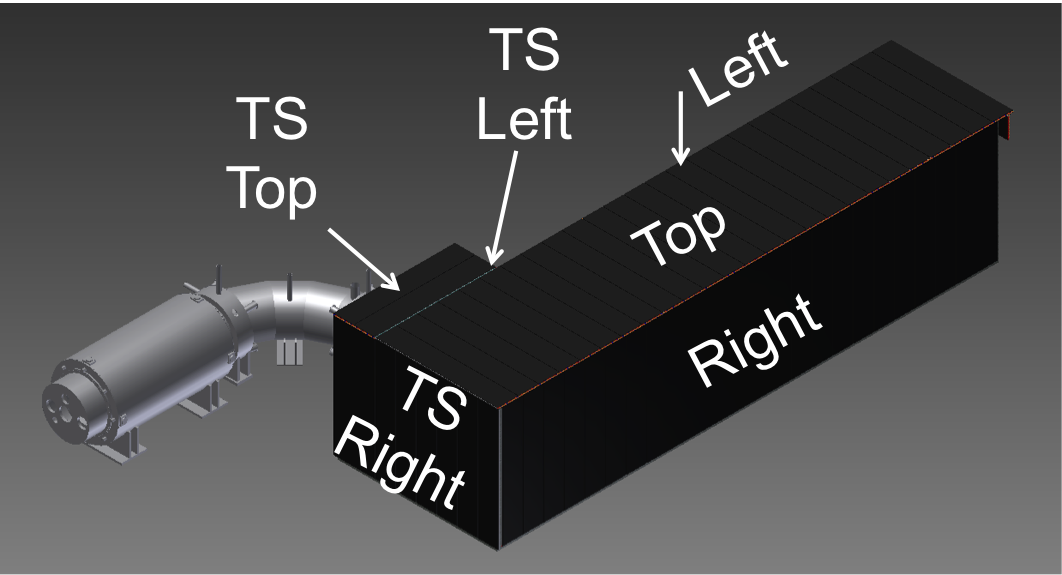}
  \includegraphics[width=2.8in]{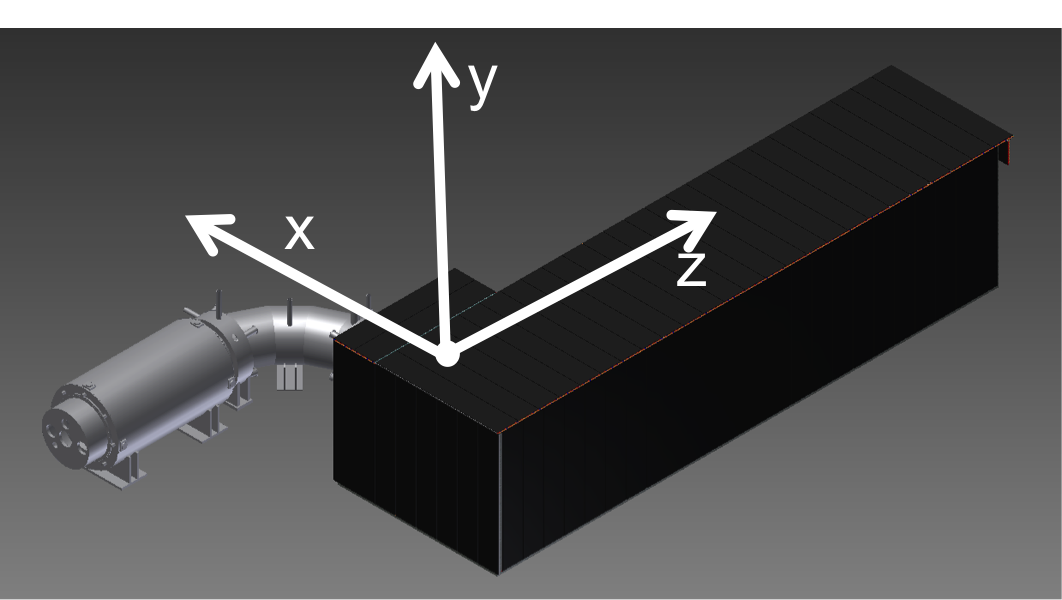}
  \caption{The layout of the Cosmic Ray Veto system (left) and shown with the coordinate system(right).}
\label{fig:CRV}
\end{figure}

The concept of the Mu2e experiment is to stop low-momentum muons from a pulsed beam on an aluminum target to form muonic atoms and then to measure the resulting electron spectrum.  The signal would produce a mono-energetic electron with energy of about 105~MeV.  In order to reach the design sensitivity (single-event sensitivity of $2.5 \times 10^{-17}$), the experiment must stop about $10^{18}$ muons.  Keeping the background expectation to less than one event in this high-intensity experiment is challenging and results in the unique experimental setup described below and shown in Fig.~\ref{fig:Mu2e}.  The cosmic-ray-veto detector (CRV) is not shown in Fig.~\ref{fig:Mu2e}, but it surrounds the Detector Solenoid and is split into several regions defined in Fig.~\ref{fig:CRV}.  This paper reviews the Mu2e experiment with a focus on sources of neutron and gamma radiation that affect the environment of the CRV.  Note that neutron and gamma radiation will even be more important for potential intensity upgrades to Mu2e that are being discussed~\cite{MU2E2}.

\section{The Mu2e Experiment}

The first step in the experiment is to produce the low-momentum pulsed muon beam.   Accelerator infrastructure from the antiproton rings used in the Tevatron program will deliver 8~GeV protons to the production target.  Located in the Production Solenoid (PS), the production target is a major source of neutron and gamma radiation. About $3\times10^7$ protons per bunch will strike the target.  Resonant extraction from a bunched beam stored in a ring with cyclotron frequency 1694 ns produces beam pulses of about 200 ns width every 1694 ns.  This pulse frequency is well suited for the Mu2e experiment given that the lifetime of the muonic aluminum is about 864~ns.   Pions
and muons produced inside the production solenoid are collected and passed to the
S-shaped muon Transport Solenoid (TS), where absorbers and collimators are optimized to stop most antiproton contamination while efficiently transmitting the negatively charged pions and muons.  The absorbers and collimators in the TS region provide a second major source of radiation.   Most pions decay inside the 13~m long
beamline. About 40\% of muons exiting the beamline will be stopped in
the aluminum target producing a rate of about $0.0016$ stopped
muons per proton.  The remaining muons will pass through the detector solenoid region into the Muon Beam Stop, a third major source of neutrons and gamma rays.

Muons stopped in the target will form a muonic atom.   About 60\% of stopped muons will undergo muon capture on the nucleus while the other 40\% will decay in orbit (DIO).  The muon capture process on aluminum is a fourth major source of radiation.  The DIO process produces an electron with a continuous Michel distribution including a long high energy tail due to radiative corrections with an endpoint energy equal to the conversion electron energy.  In order to combat the DIO background, the Mu2e experiment requires a tracking detector with momentum resolution of about 0.1\%.    

The primary backgrounds for the Mu2e experiment can be classified into several categories: prompt, muon-induced, and cosmic rays.  An example of a prompt background is $\pi^{-}$ radiative capture and
subsequent conversion of the $\gamma$ in the stopping target material,
which can produce a 105~MeV electron.  Backgrounds like this one are controlled by taking advantage of the muon lifetime and optimizing the properties of the pulsed beam.  After a beam pulse there is a delay of 670~ns, allowing sufficient time for the pions from the main proton pulse to dissipate, before the time window begins and reconstructed electrons are considered as signal candidates.  To avoid prompt backgrounds in the signal time window, Mu2e requires the fraction of protons outside the beam pulse to be less than $10^{-10}$. Beam electrons, muon decay in flight, and pion decay in flight are other prompt backgrounds that are defeated through the combination of the pulsed beam and delayed signal window.


  Cosmic-ray muons may interact within the DS region and produce background electrons.  Passive shielding and an active CRV system are employed to ensure that cosmic rays are a sub-dominant background.  The CRV surrounds the detector solenoid and the terms used to refer to different regions of the CRV are defined in Fig.~\ref{fig:CRV}.  The active shield is plastic scintillator read out by wavelength-shifting fibers and Silicon photodetectors (SiPMs).  We expect to require hits in multiple layers of the CRV as a veto requirement on cosmic ray muons.  Note that this veto will not be applied online at the trigger level. All events will be stored and the veto algorithm can be optimized offline.  A veto efficiency of 99.99\% is required of the CRV in order to ensure that the cosmic ray contribution to the background is sub-dominant.  

   The background expectation in the Mu2e experiment for a  three-year run at 8~kW beam power is about 0.5 events where the background contribution from cosmic rays is about 0.05 events.  Note that the CRV is a critical component if Mu2e is to achieve single-event sensitivity -- without the CRV there would be hundreds of background events over the life of the experiment (about one event per day).

\section{Studies of neutron and gamma radiation impacts on the CRV}

 We originally designed a three-layer CRV and planned to impose a veto defined by a time coincidence of two out of three in-line layers of scintillator~\cite{Mu2eCDR}.  False vetoes can be created from neutron or gamma radiation through accidental coincidences of hits in two layers (hereafter referred to as coincidental hits)in which a different particle hits two layers within the veto time window, or from correlated hits in which the same particle deposits energy in multiple layers of the CRV.  An upper limit on the rate of single hits in a given counter can be obtained by using the requirement that the CRV produce no more than 1\% deadtime for Mu2e, as well as the CRV geometry, expected 5 ns CRV timing resolution, and a 50 ns veto window.  The same criterion leads to limits on the correlated hit rate as well.  For the coincidental hits this calculation yields an upper limit of about $2 \times 10^8$ hits/cm$^{2}$, while for correlated hits the tolerable upper limit is about $6 \times 10^5$ hits/cm$^{2}$ over the lifetime of the experiment.  To convert these into an upper limit on the particle rate one must know the efficiency for particles to produce hits in single counters for the coincidental hits or in multiple counters for the correlated hits.

We studied the efficiency for neutrons and gammas to produce hits in our plastic scintillator counters using the G4BEAMLINE~\cite{g4bl} simulation package and the QGSP\_BERT\_HP physics list. For each particle passing through a counter we calculate the total energy deposited with corrections for Birk's law~\cite{Birks} and require the deposited energy to be above 1 MeV to be considered a hit (for reference, a minimum ionizing muon would deposit about 4 MeV in our 2 cm thick counters).  The efficiencies for neutrons to leave a hit in a counter are about 0.5\% for less than 1 eV, about 0.1\% for between 1 eV $< E <$ 2 MeV, and about 10\% for energy above 2 MeV.  For gammas below 1 MeV the efficiency is effectively zero, while above 1 MeV it is about 10\%.  For gammas we find that the efficiency for leaving a hit in at least two layers is about 1\% for gammas above 5 MeV and that this is the most important contribution to the deadtime produced by false vetoes.

\begin{figure}[hbt]
  \centering
  \includegraphics[width=2.8in]{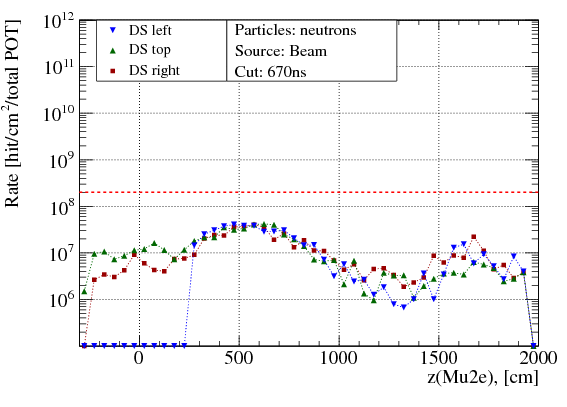}
  \includegraphics[width=2.8in]{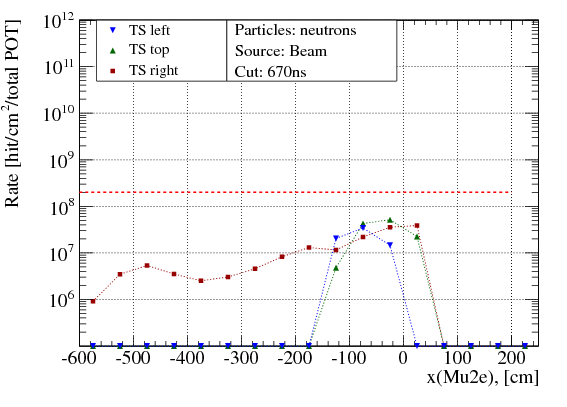}
  \includegraphics[width=2.8in]{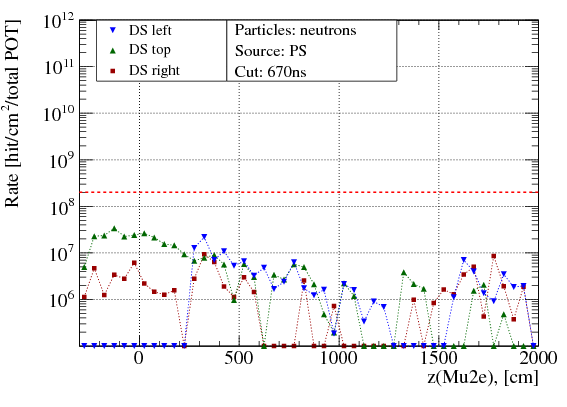}
  \includegraphics[width=2.8in]{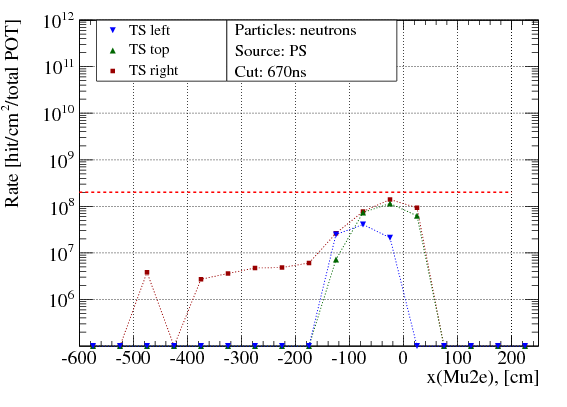}
  \caption{\label{fig:NeutronSingleRates}  The Neutron rates in the DS
    region(left) and TS region(right) of the CRV for each counter
    after applying the 670~ns timing requirement.  Beam sources are shown on the top and PS sources are shown on the bottom.  The red line shows the rate
    limit as described in the text.}
\end{figure}

\begin{figure}[hbt]
  \centering
  \includegraphics[width=2.8in]{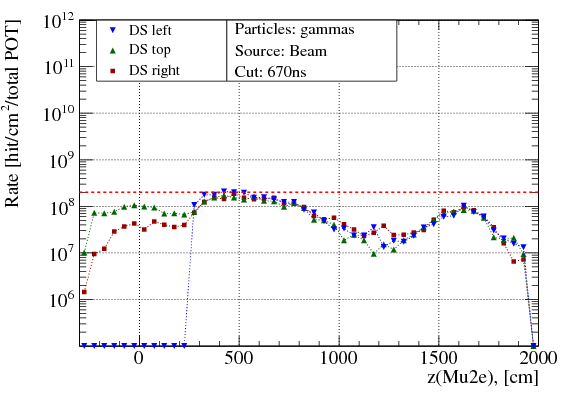}
  \includegraphics[width=2.8in]{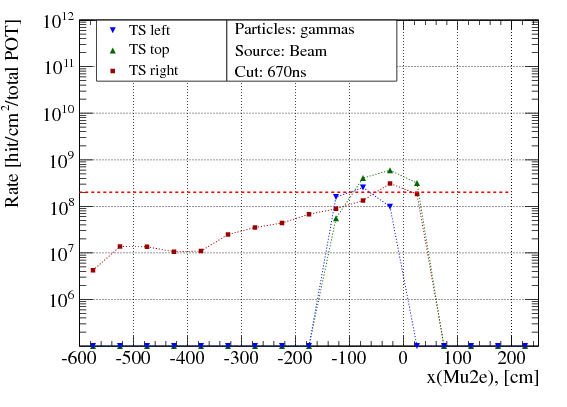}
  \includegraphics[width=2.8in]{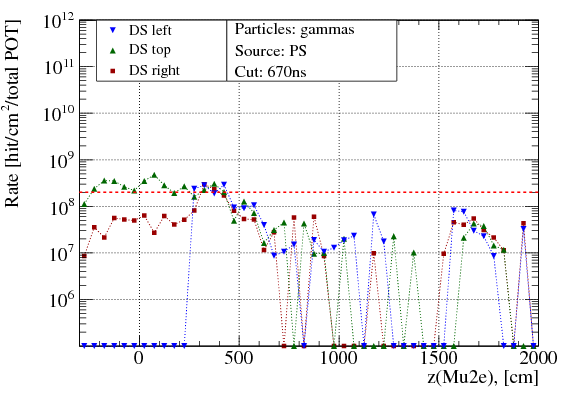}
  \includegraphics[width=2.8in]{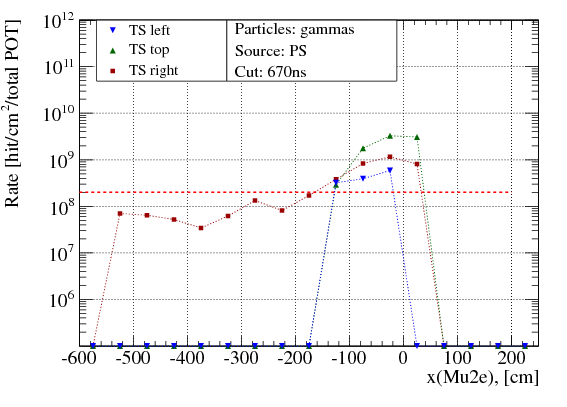}
  \caption{\label{fig:GammaSingleRates}  The Gamma rates in the DS
    region(left) and TS region(right) of the CRV for each counter
    after applying the 670~ns timing requirement.  Beam sources are shown on the top and PS sources are shown on the bottom. The red line shows the rate limit as described in the text.}
  
\end{figure}

\begin{figure}[hbt]
  \centering
  \includegraphics[width=2.8in]{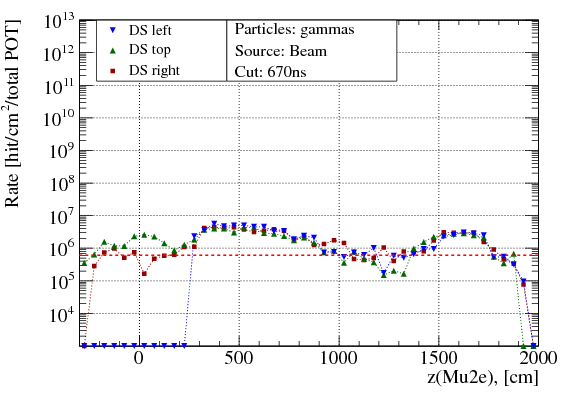}
  \includegraphics[width=2.8in]{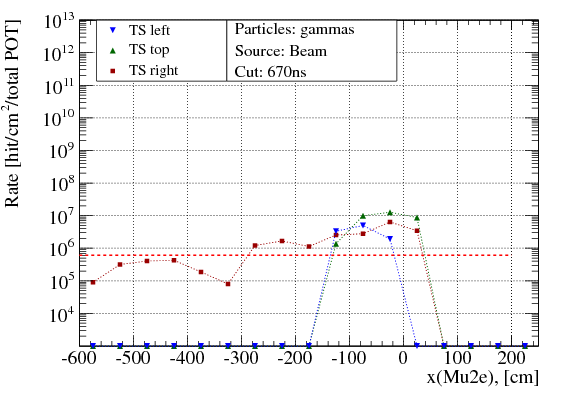}
  \includegraphics[width=2.8in]{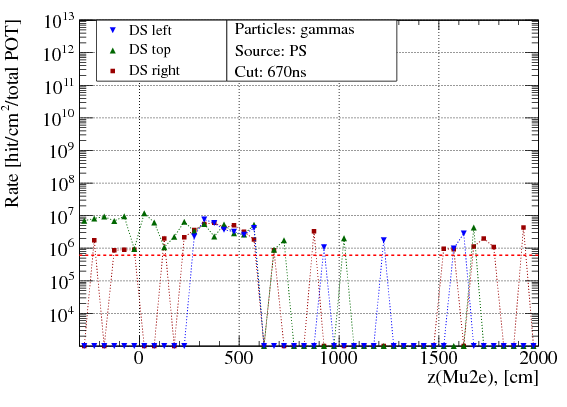}
  \includegraphics[width=2.8in]{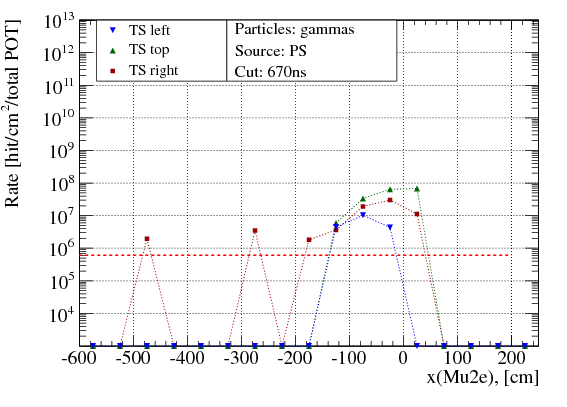}
  \caption{\label{fig:GammaDoublesRates}  The Gamma two-out-of-three hit
    rates in the DS region(left) and TS region(right) of the CRV for
    each counter after applying the 670~ns timing requirement.  Beam sources are shown on the top and PS sources are shown on the bottom.  The red line shows
    the rate limit as described in the text.}
\end{figure}

Using the full Mu2e geometry and shielding, the neutron and gamma rates and energy distributions are generated for all regions of the CRV.  The efficiencies described above are folded into these particle rates to obtain hit rates which can be compared to the upper limits derived from the deadtime requirement.  The obtained rates are shown in Figs.~\ref{fig:NeutronSingleRates}--~\ref{fig:GammaDoublesRates}.  Note that we applied the Mu2e signal timing requirement of $670$~ns$< t < 1595$~ns with respect to the proton pulse.  This removes a large contribution from fast neutrons.  In fact, with these timing cuts, neutrons are not a significant issue for the CRV.  However, we find that coincidental hits from gammas produce a rate near the limit, and that correlated hits from gammas produce a rate about one order of magnitude above the allowable rate over much of the CRV.  Through additional studies, we find that the latter is controlled by increasing the number of scintillator layers from three to four and requiring signals from three out of four layers.  Based on these studies the CRV has been redesigned to include four layers.

\section{Conclusions}

In this work, we found that the gamma rates at the CRV were too high, producing singles rates slightly above the upper limit imposed by a 1\% deadtime, but producing correlated hits about an order of magnitude above the allowable limit.  We studied several ways to modify the Mu2e shielding and the CRV design to solve this problem.  These studies provided the primary motivation to move from a three-layer to a four-layer CRV design.  The four-layer design improves that rates in the CRV significantly since it has an efficiency for gammas to produce correlated hits in three-out-of-four layers that is about two orders of magnitude lower than the two-out-of-three efficiency in the three-layer design.  With the new design and optimized shielding, we are confident that the CRV will provide a high-efficiency veto for muons in the difficult high-intensity environment of Mu2e.


\end{document}